\documentclass[twocolumn,widetext,amsmath,amssymb]{revtex4}
\usepackage{graphicx}% Include figure files
\usepackage{dcolumn}% Align table columns on decimal point
\usepackage{bm}% bold math
\begin{document}

\title{Gravity vs Radiation Models: On the Importance of Scale and Heterogeneity in Commuting Flows}

\author{A. Paolo Masucci$^1$}
\author{ Joan Serras$^1$}
\author{Anders Johansson$^{2,1}$}
\author{ Michael Batty$^1$}
\affiliation{1-Centre for Advanced Spatial Analysis, University College  London, W1T 4TJ, London, UK \\
2-Systems Centre, Department of Civil Engineering, University of Bristol,  BS8 1UB, UK}
\date{\today}

\begin{abstract}
We test the recently introduced \textit{radiation model} against the \textit{gravity model} for the system composed of England and Wales, both for commuting patterns and for public transportation flows. The analysis is performed both at macroscopic scales, i.e. at the national scale, and at microscopic scales, i.e. at the city level. It is shown that the thermodynamic limit assumption for the original radiation model significantly underestimates the commuting flows for large cities. We then generalize the radiation model, introducing the correct normalisation factor for finite systems. We show that even if the gravity model has a better overall performance the parameter-free radiation model gives competitive results, especially for large scales.
\end{abstract}

%\keywords{Suggested keywords} % Use showkeys class option if keyword
                               % display desired

\maketitle

%%%%%%%%%%%%%%%%%%%%%%%%%%%%%%%%%%%%%%%%%%%%%%%%%%%%%%%%%%%%%%%%

\medskip

\section{Introduction}

In the last year, progress has been reported on theories
providing a framework for human commuting patterns \cite{ref1_simini,ref2_noulas}. Both
papers suggest that the main ingredient in a `universal' law
predicting human mobility patterns is  topological, i.e. it
does not directly depend on metrical distance.
This discovery aims to rewrite the assumptions that have been made during the last century on mobility patterns and in particular the traditional \textit{gravity model} first suggested for use in human interaction systems by Carey (1859) \cite{carey} and popularised by Zipf in 1946 \cite{ref3_zipf}  and the \textit{intervening opportunities model} introduced by Stouffer in 1940 \cite{ref4_stouffer}.
It is worth noticing that lately, purely topological relations have also been found to be leading components for the explanation of animal collective behaviour  \cite{ref5_ballerini}.

In particular in \cite{ref1_simini} a simple theory called the
\textit{radiation model}, based on diffusion dynamics, has been
developed and the model appears to match experimental data well. The model gives exact analytical results and it has the additional desirable feature of being parameter-free, i.e. it has the characteristics of a universal theory.

In this contribution we  use three different datasets in
order to assess the universality, accuracy, and robustness of the
newly proposed radiation model applied to human mobility and public
transport infrastructure. The datasets we use are available as: (i) a complete multimodal network for transportation in the UK, comprising the road network for bus and coach, the rail networks for tube and rail, and the airline networks for plane. The weights on these networks consist of the volumes of the transport (vehicles, trains, planes) from transport time-tables; (ii)  commuting patterns for England and Wales at ward level resolution from the 2001 Population Census; and (iii)  population density data for the UK at ward level resolution, also from the Census.

Our first concern about the radiation model is the presumption of universality. In our interpretation, `universality' means that the model can be applied at all spatial scales, all time periods, and to different places. Regarding the system scale, we show that among cities, the radiation model is broadly accurate for commuting, while it is not accurate at all in forecasting both the transportation patterns between cities, or for the commuting flows within London. Regarding the applicability of the model to different countries, we notice that the radiation model is normalised to an infinite population system. We derive the correct normalisation for finite systems and we show that it deviates from the one derived in  \cite{ref1_simini} at the thermodynamic limit. This deviation is not really appreciable for large population systems at the scale of counties in the US, but it becomes relevant for smaller systems composed of much smaller but equivalent entities such as wards in the regions including England and Wales.

\subsection{The gravity model}

The gravity model is based on  empirical evidence that the
commuting between two places \textit{i} and \textit{j}, with
origin population $m_i$ and destination population $n_j$, is proportional to the
product of these populations and inversely proportional to a power law
of the distance between them. Many studies have been carried on such a model, where it is often subject to additional constraints on the generation and attractions of flows, and on the total travel distance (or cost) observed. These variants can be derived consistently using information minimising or entropy maximising procedures  \cite{ref6_wilson}.

In our research we employ two models. One is a four-parameter one,
that is the one also used in \cite{ref1_simini} and was first stated in this form by Alonso (1976) \cite{alonso}:

\begin{equation}
G_{ij} =A\frac{m_{i}^{\alpha } n_{j}^{\beta } }{r^{\gamma } },
\label{eq_1_}
\end{equation}
where \textit{A} is a normalisation factor and $\alpha $,
$\beta $ and $\gamma $ are the parameters of the model, which can be
determined by multiple regression analysis.

The second is a simpler perhaps more elegant model, that just carries the
parameter as the exponent of the denominator and it is the one that is more frequently used in transportation modelling
\begin{equation}
G_{ij} =A\frac{m_{i} n_{j} }{r^{\gamma } }.
\label{eq_2_}
\end{equation}

\begin{figure} \begin{center}
	\includegraphics[width=0.5\textwidth]{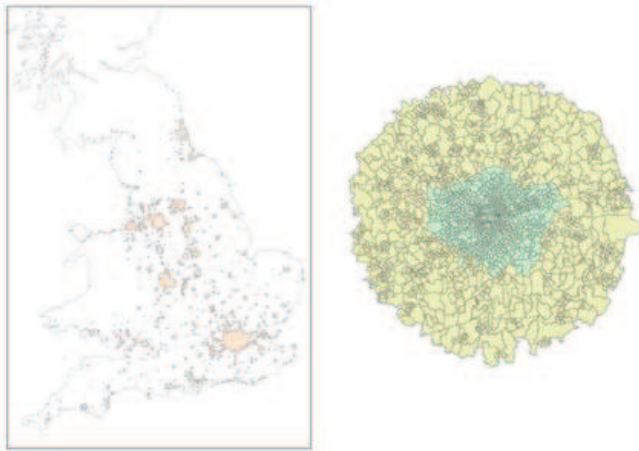}
	\caption{(Colour online) The geographical areas considered in the present analysis. Left panel: Cities of England and Wales. Right panel: wards of the GLA (Greater London Authority) and the surrounding Outer Metropolitan Region.}
	\label{fig1}
\end{center}
\end{figure}

\subsection{The radiation model}

The radiation model tracks its origin from a simple particle
diffusion model, where particles are emitted at a given location and
have a certain probability \textit{p} of being absorbed by surrounding
locations. It comes out that the probability for a particle to be
absorbed is independent of \textit{p}, but it depends only on the
origin population $m_{i}$, the destination population $n_{j}$ and
on the population in a circle whose centre is the origin and radius
the distance between the origin and the destination, minus the
population at the origin and the population at the destination,
$s_{ij}$. Then the number of commuters, that we call $T_{ij}
^{\infty }$, from location \textit{i} to location \textit{j} is
estimated to be a fraction of the commuters from population
\textit{i}, $T_{i} $, that is:

\begin{equation}
T_{ij} ^{\infty } =T_{i} \frac{m_{i} n_{j} }{(m_{i} +s_{ij} )(m_{i} +n_{j} +s_{ij} )}.
\label{eq_3_}
\end{equation}

\begin{figure} \begin{center}
	\includegraphics[width=0.5\textwidth]{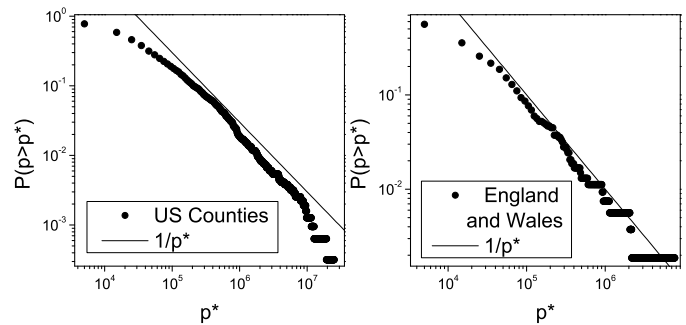}
	\caption{In the left panel, the cumulative frequency
distribution for the population size \textit{P(p$>$p*)} for the
US counties analysed in \cite{ref1_simini} . In the right panel, the cumulative
frequency distribution for the population size \textit{P(p$>$p*)
}for the cities of England and Wales.}
	\label{fig2}
\end{center}
\end{figure}

The most interesting aspect of Eq. \eqref{eq_3_} is
that it is independent of the distance and that it is parameter
free. Nevertheless Eq. \eqref{eq_3_} has been derived in the
thermodynamic limit, that is for an infinite system. It is easy to
show that for a finite system the normalisation brings us to a slightly
different form of the radiation model, that is

\begin{equation}
T_{ij}^{} =\frac{T_{ij} ^{\infty } }{1-\frac{m_{i} }{M } } =\frac{T_{i} ^{} }{1-\frac{m_{i} }{M } } \frac{m_{i} n_{j} }{(m_{i} +s_{ij} )(m_{i} +n_{j} +s_{ij} )},
\label{eq_4_}
\end{equation}
where $M =\sum _{i}m_i  $ is the total sample
population and we have $T_{ij}^{} \to T_{ij} ^{\infty } $ for $M
\to \infty $.

In a finite system $T_{ij} ^{\infty } $ underestimates the
commuting flows by a factor $\frac{1}{1-\frac{m_{i} }{M } } $.
For a very large system with uniform population Eq.~\eqref{eq_3_} is a very good approximation, but actually the city size distribution is not uniform for it usually follows a very heterogeneous skewed distribution, such as Zipf's law  \cite{ref7_zipf49}.

To understand the deviations of Eq. \eqref{eq_3_}
from Eq. \eqref{eq_4_}, we measure the factor
$F=\frac{1}{1-\frac{m_{i} }{M } } -1$ for the dataset used in
\cite{ref1_simini} and for a smaller system: the region composed of England and Wales. In the former case the US
system  is very large. The analysis is performed at the
county level and that reduces the population heterogeneity of the
system. We find that the largest deviation is in the flows from
Anderson county and this is of around $F\approx 6\% $. This is not a
particularly large deviation, but the same measure for England and Wales for example brings a deviation
for the commuting flows from London $F\approx 17\% $, that is a considerably
larger deviation.

As we have shown that Eq. \eqref{eq_3_} is not
universal, but scale dependent, a better choice for our
investigation of UK commuting patterns is Eq. \eqref{eq_4_}.

In \cite{ref1_simini} $T_i$ is considered to be proportional to $m_i$,
that is a good estimate, while in our analysis we derive its value
directly from the commuting network $w_{ij}$, i.e. $T_i=\sum_j{w_{ij}}$, that is in network theory terminology the
out-strength of location \textit{i} \cite{dorog}.

Moreover in \cite{ref1_simini} the model is based on job opportunities that are
considered to be proportional to population.

In fact Eq. \eqref{eq_4_} can be rewritten in network theory terminology.
Hence given that $w_{ij}$ are the elements of the weighted
directional adjacency matrix representing the commuting between
locations \textit{i} and \textit{j}, we define the out-strength
of vertex \textit{i} as $s_{i}^{out} =\sum_{i}w_{ij}$, and the in-strength as $s_{i}^{in} =\sum_{j}w_{ij}$.
Then we have
\begin{equation}
w_{ij} =\frac{s_{i}^{out} }{1-\frac{s_{i}^{in} }{\sum _{ij}w_{ij}  } } \frac{s_{i}^{in} s_{j}^{in} }{(s_{i}^{in} +S_{ij} )(s_{i}^{in} +s_{j}^{in} +S_{ij} )}.
\label{eq_5_}
\end{equation}
Here $S_{ij} =\sum _{k\in K_{ij} }s_{k}^{in}$ and $K_{ij}=\left\{\forall k:d(i,k)>0,d(i,k)<d(i,j)\right\}$ where
\textit{d(i,j)} is the distance between \textit{i} and \textit{j}.
 Eq. \eqref{eq_5_} is an interesting relation between the commuting flows of the network which can be verified in itself. In fact the in-strength of a given vertex represents the job opportunities in that location, since it quantifies exactly the number of people going to work in that location.

\section{Data Analysis}

In this section we test the models defined in Eq.
\eqref{eq_1_}, Eq. \eqref{eq_2_} and Eq.
\eqref{eq_4_} against empirical data. In the first subsection
we analyse the commuting between the cities of England and Wales
(see left panel of Fig.~\ref{fig1}), thereby simulating  the models at
macro-scales, while in the second subsection we analyse the
commuting between the wards of London (see right  panel of Fig.~\ref{fig1}),
simulating  the models at micro-scales.

\begin{figure*} \begin{center}
	\includegraphics[width=0.7\textwidth]{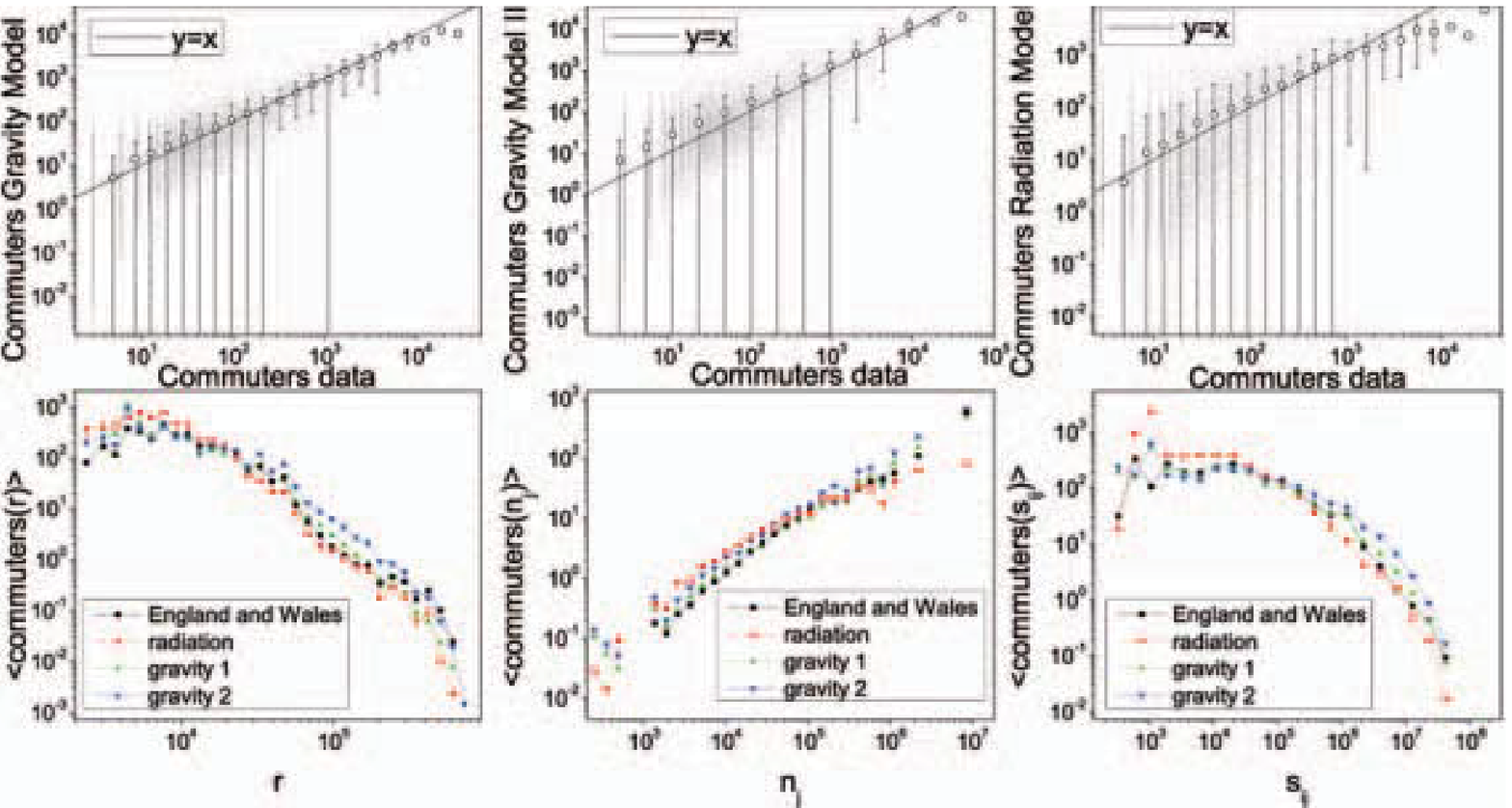}
	\includegraphics[width=0.7\textwidth]{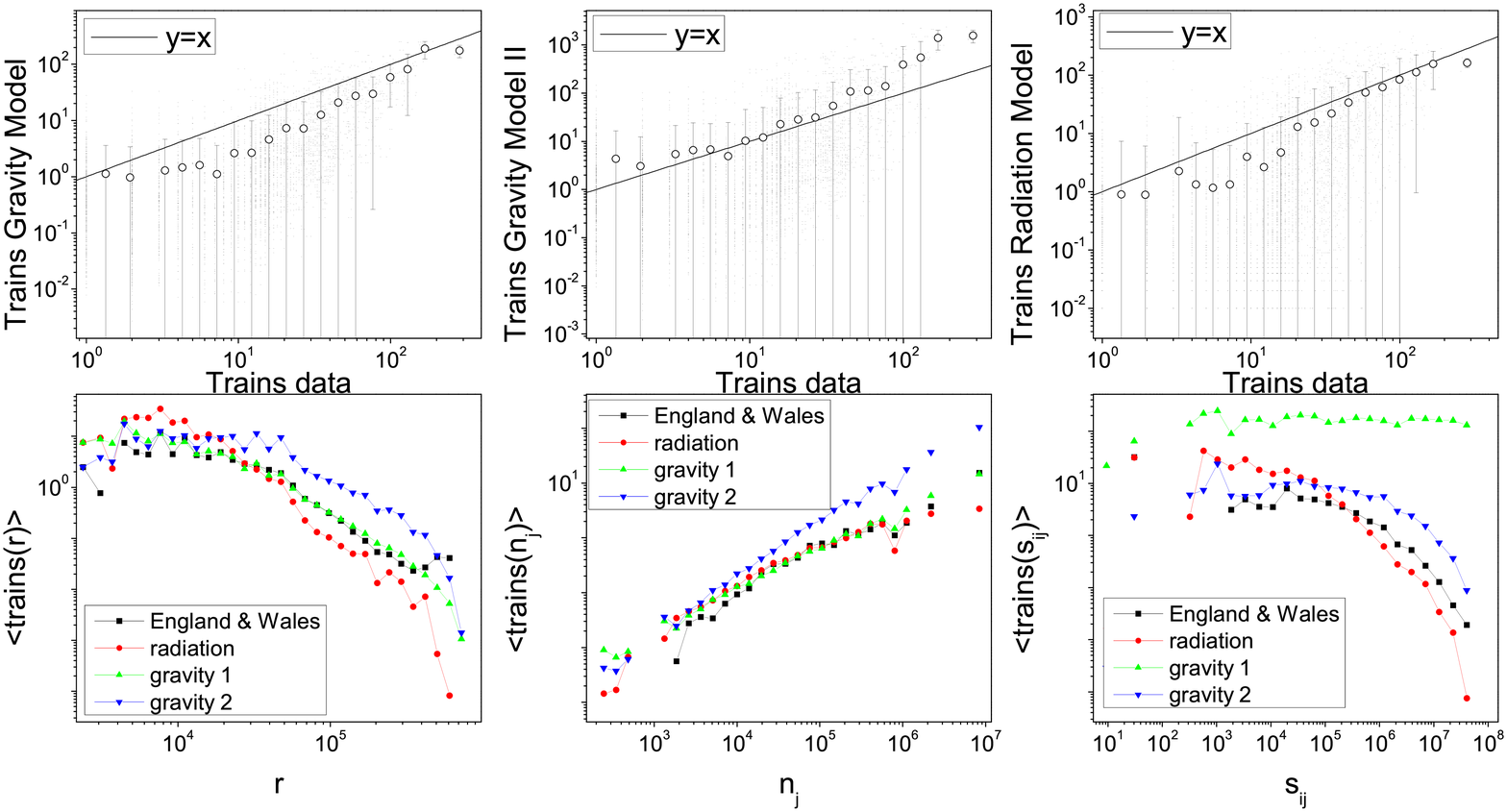}
	\includegraphics[width=0.7\textwidth]{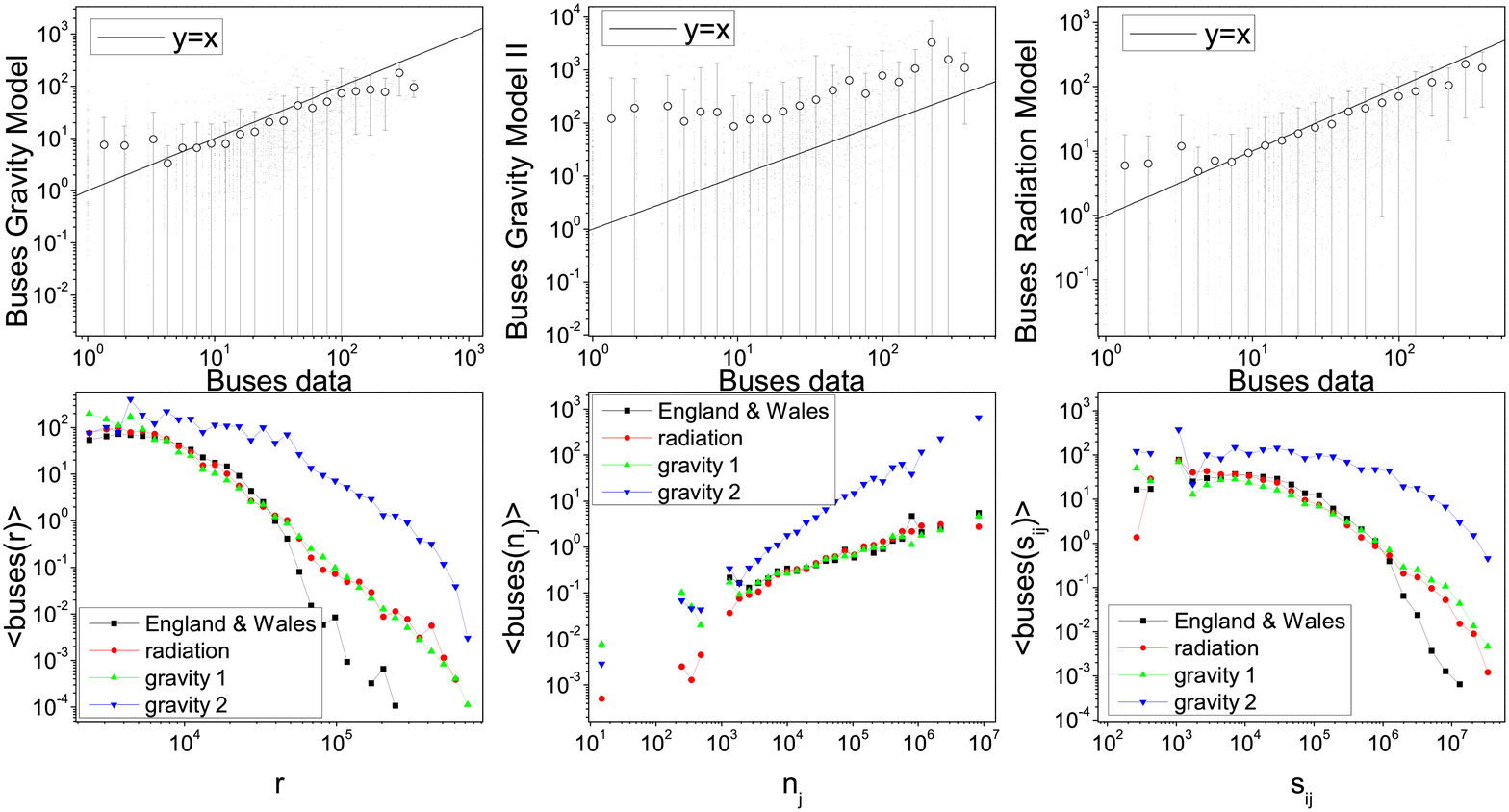}
	\caption{(Colour online) Analysis results using England and Wales city clusters.
	Top panel: Census 2001 commuter flows (parameters for the gravity model: $\alpha=0.81$, $\beta=1.03$, $\gamma=2.42$. Parameters for the gravity model II: $\gamma=1.54$). Middle panel:  Trains flows (parameters for the gravity model: $\alpha=0.76$, $\beta=0.76$, $\gamma=1.67$. Parameters for the gravity model II: $\gamma=1.44$).
Bottom panel:  Bus flows (parameters for the gravity model: $\alpha=0.61$, $\beta=0.61$, $\gamma=2.63$. Parameters for the gravity model II: $\gamma=1.91$).}
	\label{fig3}
\end{center}
\end{figure*}

\subsection{Macroscopic Analysis: England and Wales}

In this subsection, we test the gravity model defined in Eq.
\eqref{eq_1_} and Eq. \eqref{eq_2_} and the radiation
model defined in Eq. \eqref{eq_4_} against the empirical data
for the cities of England and Wales. In this study, city clusters
have been defined via a two step process using the population for
the 8850 Census Area Statistics (CAS) of wards in England and Wales
from the 2001 Population Census  \cite{ref8_web}. In the first step, those wards with
population density above 14 persons per hectare are selected from
the rest; in the second step, adjacent selected wards have been
grouped to form a total of 535 city clusters \cite{elsa}. We show these cities
in the left panel of Fig.~\ref{fig1}.

On the top of this socio-geographical dataset, we analyse data for
the commuting between these cities using the 2001 Census  Journey to
Work data which specifies, for all surveyed commuters, the origin ward they
travel from -- their home location -- and their destination ward -- their work
location. From this data, we have calculated the number of
commuters between all pairs of cities in England and Wales.

For this study, we have also used data for the number of trains and buses
moving  between these cities. This information has been derived
from timetable data held by the National Public Transport Data
Repository (NPTDR). This data includes all public transport services
running in England, Wales and Scotland between the 5th and the 11th
of October 2009. The data is composed by two data sets: the NaPTAN
(National Public Transport Access Nodes) dataset and the
TransXChange files. The former includes all public transport nodes
categorised by travel mode  and geo-located in space. The latter one
has a series of transport modal files for each county within
England, Wales and Scotland (143 counties in total), with
information on all services running within the county. The travel
modes included are air, train, bus, coach, metro and ferry. Each
service includes routing information as a series of NaPTAN
referenced stops each with its corresponding departure and waiting
time. In
this paper, we have deduced the number of trains and buses operating
between all pairs of cities on a typical working day -- 24h -- by first
assigning a ward area to each bus and train stop via spatial
point-in-polygon queries, and then extending this assignment to city
areas.

It is worth noticing that in \cite{ref1_simini} the analysis has been made over US
counties, that are artificial units, while in this analysis we
consider cities as natural entities for  commuting. The different
choice is not merely speculative, since counties have different
physics and statistical properties than cities. It is well known
that the city size distribution follows  Zipf's law \cite{ref7_zipf49}. That means
that city size distribution has a fat tail characterised by the
scale of very large cities. The representation of the system in
terms of counties introduces an artificial cut-off in the tail of the distribution,
cutting  down the  tail, as we show in Fig.\ref{fig2}.
 It is sufficient to think of the fact
that New York City is made up of 5 different counties (boroughs), so that in a
county level analysis its population is split between those 5
counties.

\medskip

\begin{tabular}{|p{1.5in}|p{0.6in}|p{0.6in}|p{0.6in}|}
 \hline $R^2$ & Gravity I & Gravity II & Radiation \\
\hline Commuters & 0.67 & 0.00 & 0.36 \\
\hline Trains & 0.39 & 0.00 & 0.00 \\
\hline Buses & 0.11 & 0.00 & 0.32 \\ \hline
\end{tabular}

\textbf{Tab. 1:} $R^2$ calculated for the different models for England
and Wales.

\medskip

In the top panels of Fig.~\ref{fig3}, we show the analysis for the flows of commuters
in England and Wales.
 In these top panels,	 we show the comparison between the real data (x axis) and the data elaborated by
the model (y axis).
On top of that we show the average value (circles), with standard deviation bars and the line \textit{y=x},
that shows where the model meets the real data.
The gravity model parameters are estimated via  multiple regression analysis.
Further details are given in the next section, but for now we can
say that, based on $R^2$ estimates, the gravity model of Eq. \eqref{eq_1_} performs
better, followed by the radiation model, while the gravity model of Eq.\ref{eq_2_} has $R^2=0$  (see Tab. 1 for $R^2$
values).

In the following panels of Fig.~\ref{fig3}, we can see the
correlations of the commuting flows with three sensitive quantities,
the distance \textit{r} in the left panel, the destination
population $n_j$ in the central panel, the population in the
circle centred on the origin population,  with radius \textit{r},
$s_{ij}$.
 Commuter flows are strongly correlated with all these quantities.
All the plots are in log-log scale, so that apparently the correlations are in
form of power laws.

Both the radiation and the gravity model perform well in reproducing these correlations while the  gravity model II fails to reproduce the correlations with $s_{ij}$.
Also we can see that on average both the radiation and the gravity model catch the real data, so that the main difference between the two resides in the fluctuations that will be discussed in the next section.
Nevertheless, comparing the model results, we have to keep in mind
that the gravity model of Eq. \eqref{eq_1_} has 3 independent
parameters plus a normalisation factor. In the light of this, we have
to consider the parameter free radiation model as performing quite competently,
 even if the $R^2$ is not that high.

In the middle panels of Fig.~\ref{fig3}, we show the analysis for the flows of trains in
England and Wales. In this case the model that performs better is the gravity model but its performance is not satisfactory anyway (see Tab. 1).

In the bottom panels of Fig.~\ref{fig3}, we show the analysis for the flows of buses in
England and Wales.
In this case, interestingly enough the radiation model outperforms the gravity model, even if the overall result is of a generally quite poor performance, while the gravity model II again has $R^2=0$.
From the correlation analysis we can see that both the models can grab the average behaviour of the bus transportation system, while they fail to reproduce the distance and the $s_{ij}$ correlations for very large scales. Nevertheless the poor results are again given by the difficulty in reproducing the large fluctuations.

%\begin{figure} \begin{center}
%	\includegraphics[width=0.5\textwidth]{avrelerr.eps}
%	\caption{(Colour online) Analysis results using England and Wales city clusters.
%	Left panel: average relative error calculated for the commuter flows for the radiation and the gravity model. Right panel: probability density distribution for the magnitude of commuting flows.}
%	\label{fig7}
%\end{center}
%\end{figure}

\begin{figure} \begin{center}
	\includegraphics[width=0.5\textwidth]{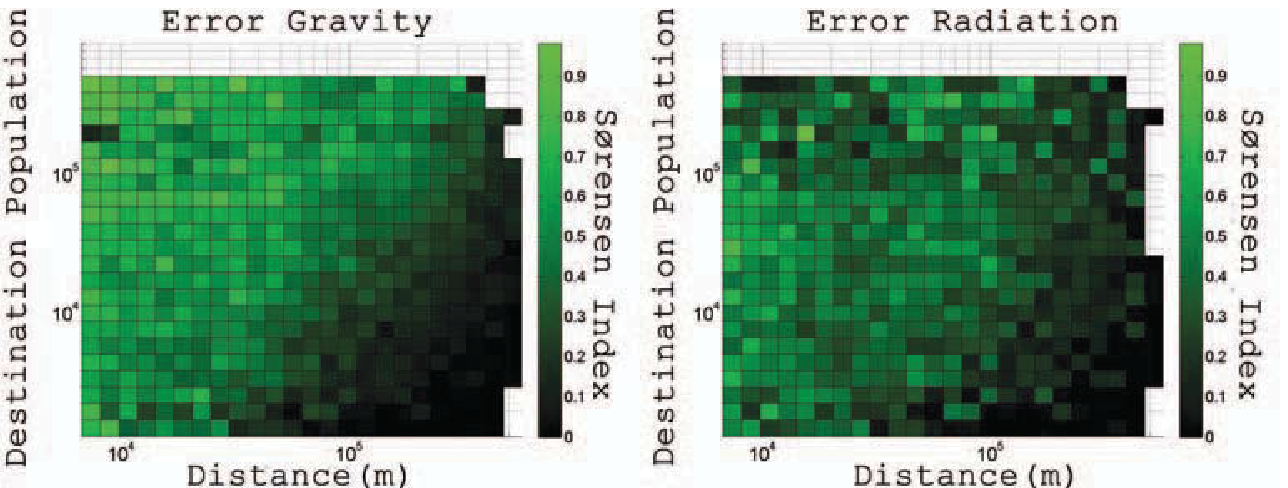}
	\includegraphics[width=0.3\textwidth]{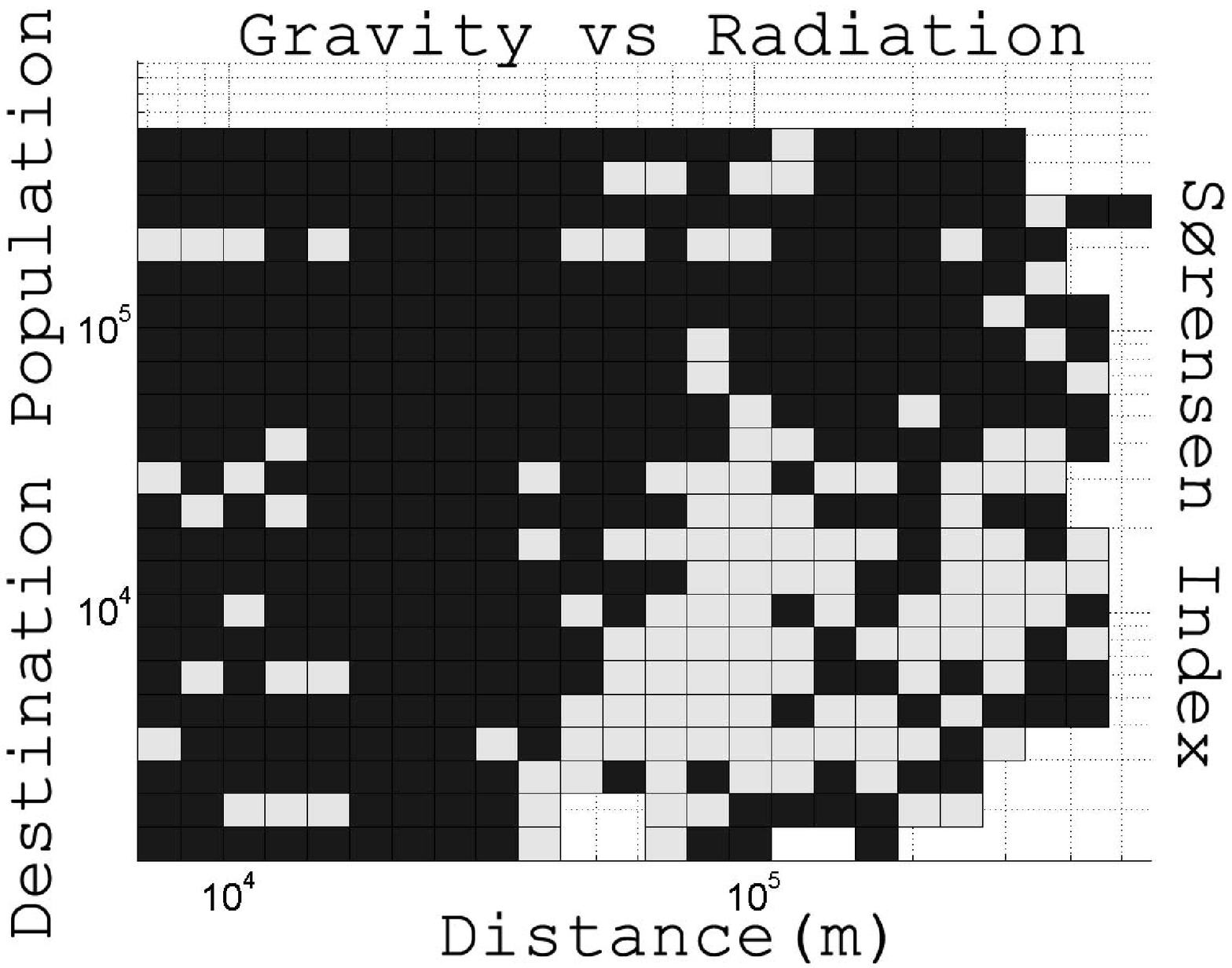}
	\caption{(Colour online) Error of flow estimates compared to
empirical  commuter flows for cities of England and Wales as a function of distance and population at destination.
In the  top panels, we show the  S{\o}rensen-Dice coefficient of Eq.\ref{eq_7_} for the gravity (left) and the radiation model (right).
The bottom panel shows areas where the gravity model performs better (black) and areas where the radiation model
performs better (light gray) using the S{\o}rensen-Dice coefficient,$E^{s{\o}rensen}$. Wherever the two models perform equally well, there is a gap in the plot (shown as white cells without borders).}
	\label{fig8}
\end{center}
\end{figure}

\subsubsection{Fluctuation analysis}

In the previous section we saw that the analysed commuting models produce reasonable $R^2$ values just for the commuting case in UK and Wales for the gravity model and the radiation model. Qualitatively, we see from Fig.\ref{fig3} that on average the models catch the real data behaviour and the sensitive parameter correlations in a striking manner. Nevertheless the $R^2$ values are not that good, especially in the case of the radiation model. To understand this, we perform a fluctuations analysis, based on the S{\o}rensen-Dice coefficient\cite{sorensen,dice}:

\begin{equation}
E^{ s{\o}rensen}\equiv\frac{2 \sum_{i,j}{ min\big(T_{ij}^{model}, T_{ij}^{empirical}\big) }}{\sum_{i,j}{T_{ij}^{empirical}} + \sum_{i,j}{T_{ij}^{model}}}
\label{eq_7_}
\end{equation}

$E^{ s{\o}rensen}$ is a similarity index that ranges from 0 to 1, where it is 0 when there is no match between  empirical and modelled data, 1 when there is a complete match.

%In the left panel of Fig.\ref{fig7}, we show the average relative error $<E^{ relative}>$ for the radiation and the gravity model as a function of the commuting flow magnitude $T_{ij}$.
%It appears clear from this figure how the fluctuations for the radiation model are much larger than the ones for the gravity model.

 % On the contrary $<E^{relative}>$ decreases monotonically for increasing flows in the case of the radiation model.

In Fig.\ref{fig8}, we show the error analysis as a function of two sensitive parameters, the distance and the destination population. In the top panels of Fig.\ref{fig8}, we show the S{\o}rensen-Dice coefficient $E^{ s{\o}rensen}$ in different locations in the phase-space made up by distance, populations at destination, and empirical flows. It is possible to see how the gravity model performs quite well for short and moderate distances, while it overestimates the flows for distances larger then 100km. On the other hand, we see how the radiation model underestimates the flows over the entire phase space.
In the bottom panel of Fig.\ref{fig8}, we show the comparison of the two models within the same phase-space for the  S{\o}rensen-Dice coefficient $E^{ s{\o}rensen}$, where the phase-space is black when the gravity model performs better than the radiation model and it is gray otherwise. From this panel it is possible to see how the gravity model performs better for short and moderate distances, where the majority of flows are concentrated, while the radiation model can better predict the commuting flows for very large distances with small and moderate population at destination.

% In particular, we see that for the gravity model the relative error is contained for flows that are smaller than 10000 commuters. On the right panel of the same figure, we show the probability density distribution for the commuting flow magnitudes in England and Wales, $P(T_{ij})$.
% From this figure the results clearly show that the great majority of commuting flows are smaller than 10000 commuters.
\begin{figure*} \begin{center}
	\includegraphics[width=0.7\textwidth]{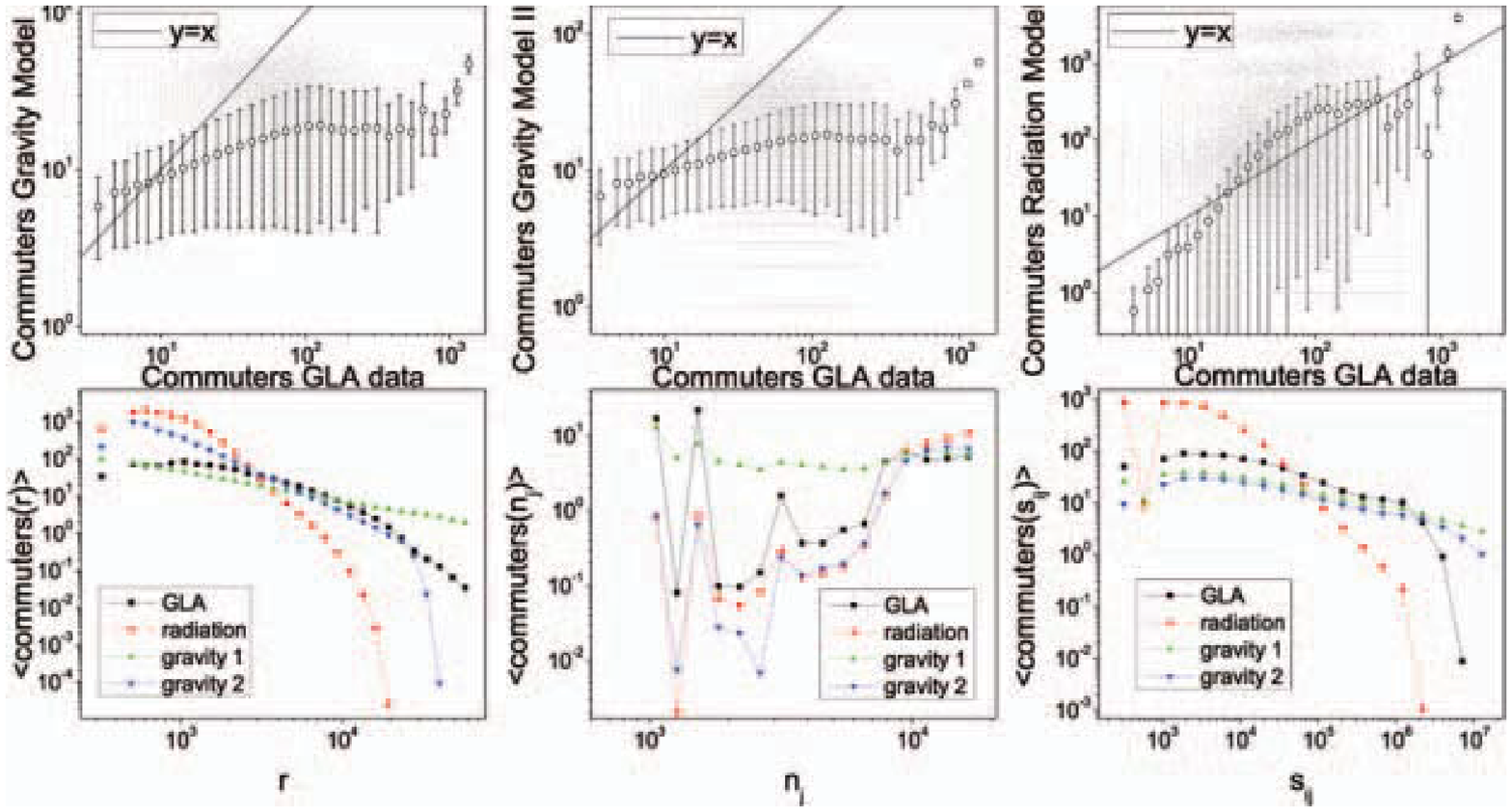}
	\includegraphics[width=0.7\textwidth]{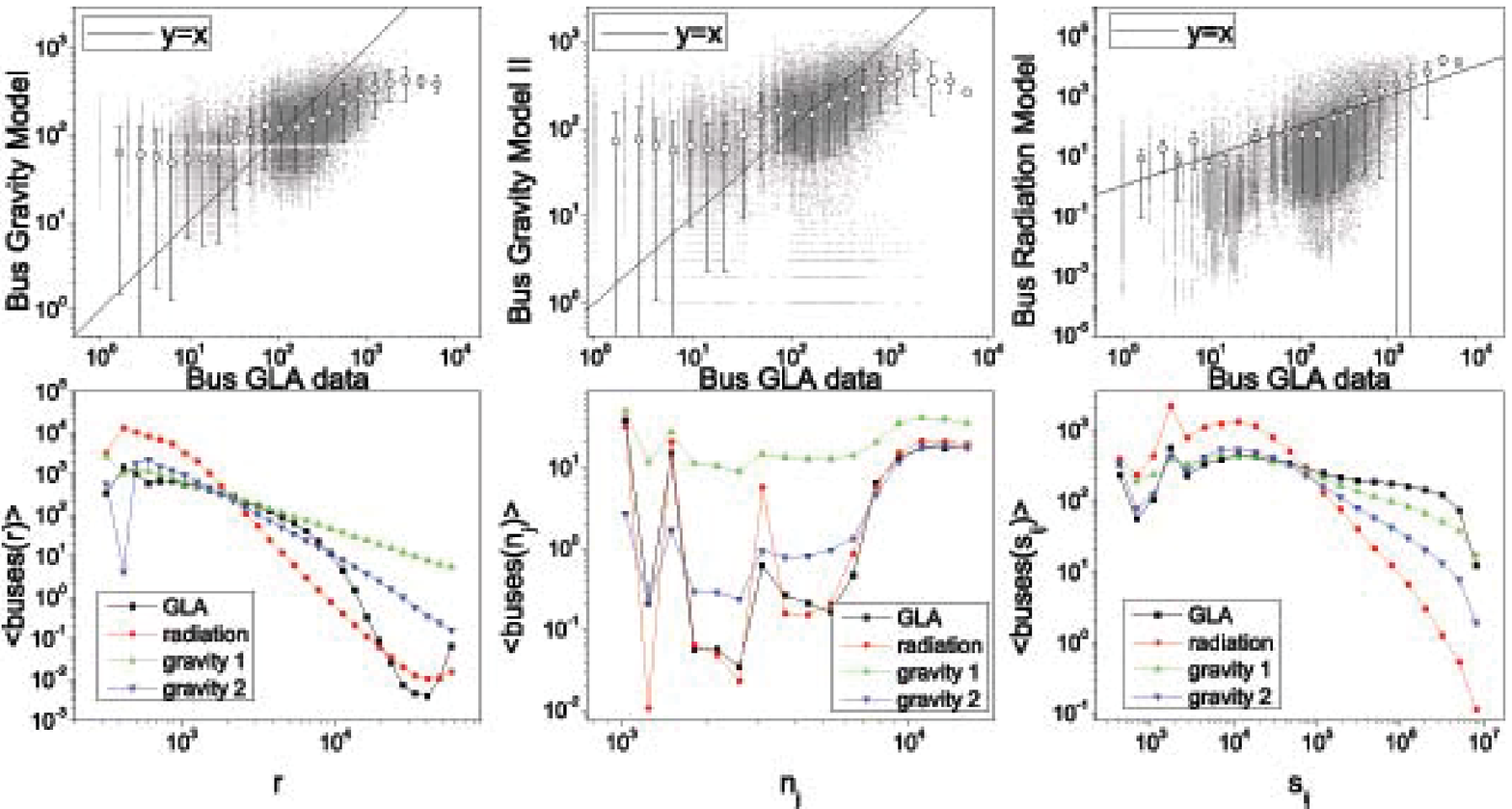}
		\caption{(Colour online) Top panels: Commuter flows analysis for GLA (parameters for the gravity model: $\alpha=0.45$, $\beta=-0.21$, $\gamma=0.82$. Parameters for the gravity model II: $\gamma=1.14$). Bottom Panels:  Bus flows analysis for GLA (parameters for the gravity model: $\alpha=0.06$, $\beta=0.09$, $\gamma=1.20$. Parameters for the gravity model II: $\gamma=2.39$).}
	\label{fig6}
\end{center}
\end{figure*}

\subsection{Small Scale Analysis: London}

In this subsection we perform the analysis on the extended Greater London Authority area at ward level. In order to do so we consider all the wards in GLA (Greater London Authority), plus the wards in the outer metropolitan area of the GLA for a radius of approximately 60km  that accounts for  commuting inside of GLA (see Fig. 1, left panel).

Our dataset, as already discussed in the previous subsection, is derived from the 2001 Census of Population with the Journey to Work data also from this Census; it gives the ward population and ward to ward commuting flows. Moreover we test the models also with the vehicular transportation, considering the number of buses travelling from ward to ward which have already been calculated (see previous subsection).

\medskip

\begin{tabular}{|p{1.5in}|p{0.6in}|p{0.6in}|p{0.6in}|} \hline
$R^2$ & Gravity I & Gravity II & Radiation \\
\hline Commuters & 0.07 & 0.06 & 0.00 \\
\hline Buses & 0.22 & 0.21 & 0.00 \\ \hline
\end{tabular}

\textbf{Tab. 2:} $R^2$ calculated for the different models for London.

\medskip

In Tab. 2 we show the results of the $R^2$ test for the
different models. We can observe straight away that the models all perform rather badly, implying that the structure of a metropolis is more complex than the one forecasted by both the radiation and by gravity models.

In the top panels of Fig.~\ref{fig6} we show the analysis for the commuting patterns, i.e. the
models against the real data. We perform a multiple regression analysis to find the best fit with the
data for Eq. \eqref{eq_1_}, whose results are shown in the figure caption.

In the second from top, left panel of Fig.~\ref{fig6}, we show  the average
number of commuters in London as a function of the  distance.
The plot shows that real data decay faster than a power law with the distance, and this behaviour is captured by none of the gravity models, that tend to follow a power law behaviour.
On the other hand the radiation model forecasts a good amount of commuting for short distances and a rapid decay, but this does not reproduce  the data well either.

In the adjacent panel we show the correlations between the commuting flows and the destination population.
  For London, this is counter intuitive, since the correlation analysis shows a few large peaks for wards with very small population. This phenomena resides in the fact that the wards where most of the jobs are concentrated in London are not residential wards. This evidence would let us think that the approximation \textit{ward population}/\textit{ward employment}  is not valid for London and that we should take this bias into account in our analysis. 

In the right panel of the same figure we show the correlations of
the number of commuters and $s_{ij}$. There are hints of a strong dependency of the commuting flows from this quantity, even if this dependency is weaker than the one reproduced by the radiation model.

\begin{figure} \begin{center}
	\includegraphics[width=0.5\textwidth]{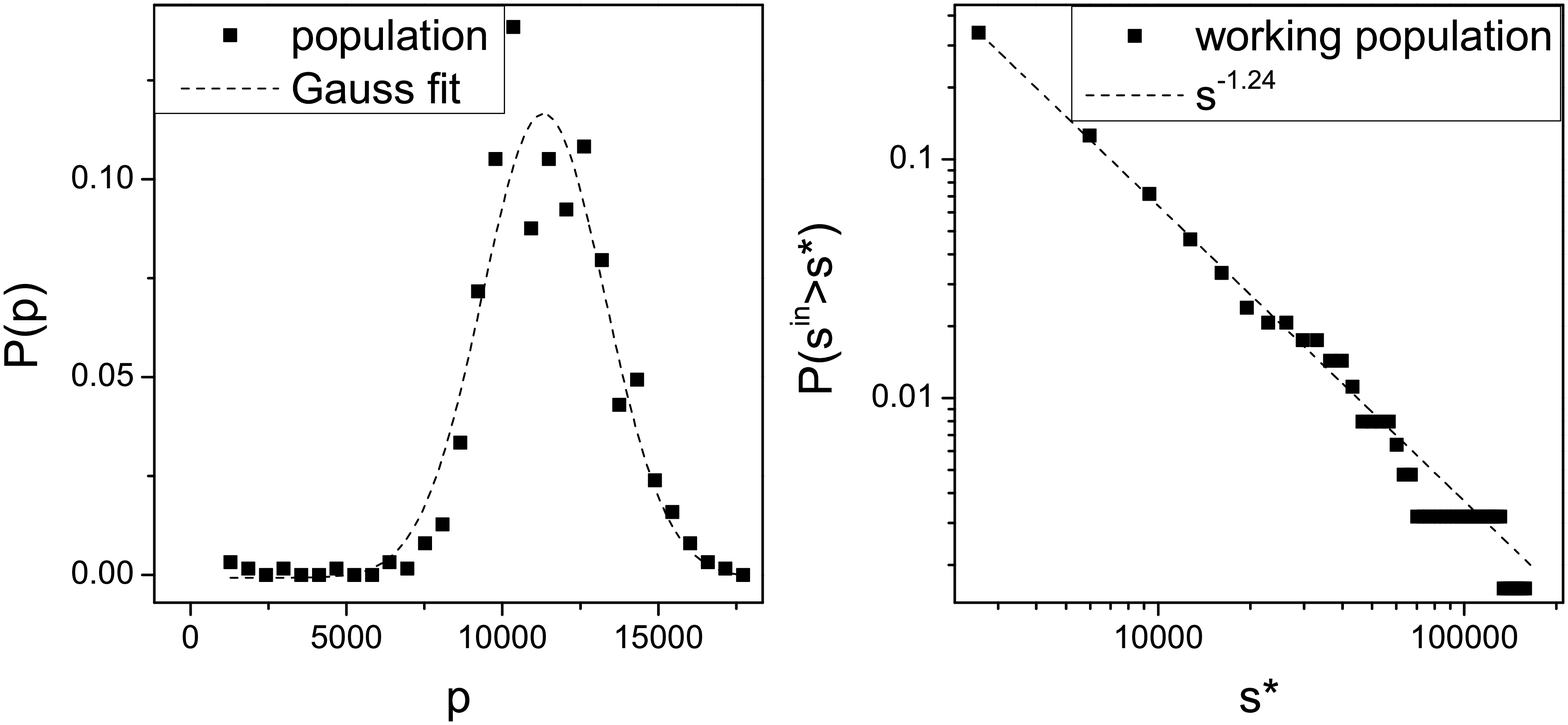}
	\includegraphics[width=0.5\textwidth]{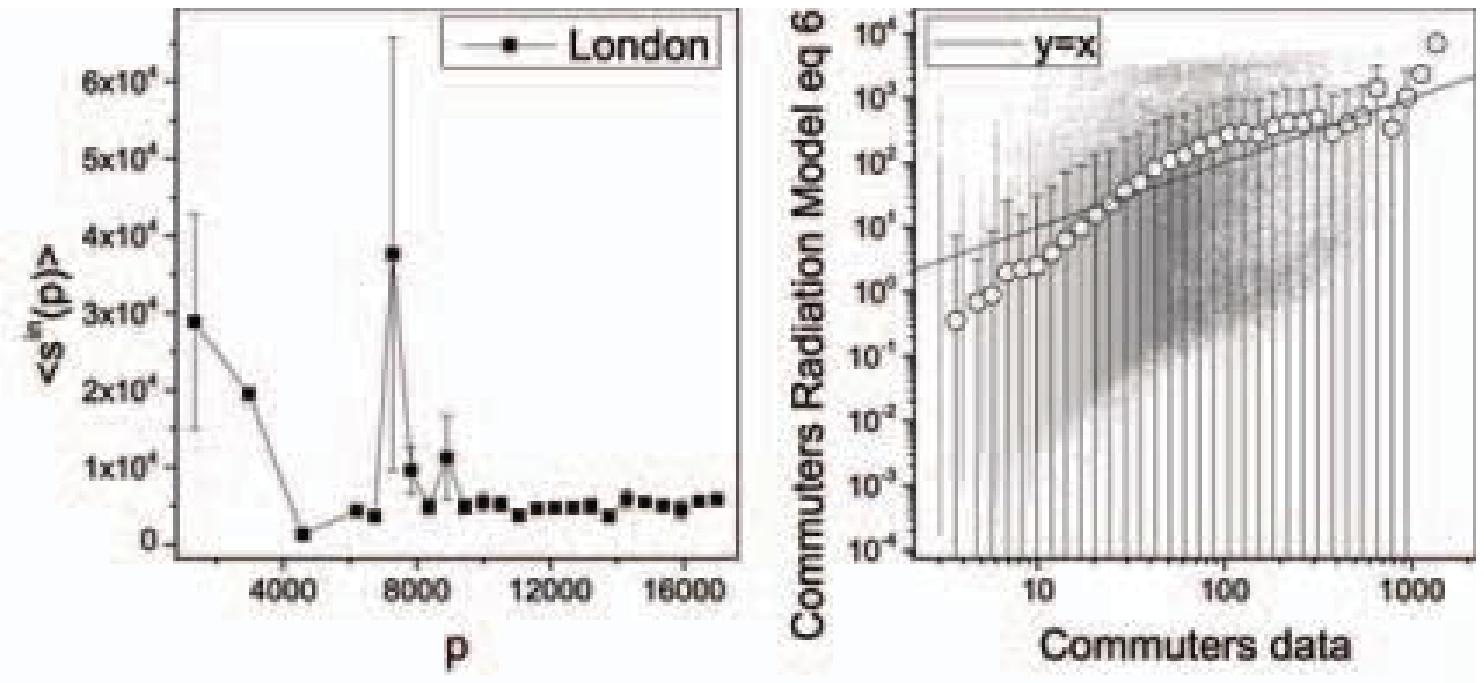}
	\caption{Top-left panel: frequency distribution of wards population size $P(p)$ in GLA. Top-right panel: cumulative frequency distribution for the working population $P(s^{in}>s^*)$  in a given ward.
	Bottom left panel: average number of employees $<s^{in}(p)>$ as a function of the ward population size. Bottom right panel: commuters flow analysis for GLA, in this case the flows are modelled via Eq.5.
	}
\label{fig12}
\end{center}
\end{figure}

In the bottom panels of Fig.~\ref{fig6}, we show the results for the analysis on the bus flows in
GLA. In Tab.~2 the $R^2$ values are displayed and we can see
that the models do not perform very well, but still better than for the
commuters case. The correlations for the number of buses with the
distance display an exponential tail, that has not been picked up by any of the models.
 As for the commuter case, we see the strongest correlations with distance and
$s_{ij}$, while the correlations with destination populations are
ill-defined.

One could argue that the poor results obtained applying the commuting models to the London intra-commuting flows could reside in the approximation validity of the employment data with the population size, in the case of the London wards.

To address this question, in the top left panel of Fig.~\ref{fig12}, we show the frequency distribution $P(p)$ of the population size $p$ for the London's wards. This is well fitted by a Gaussian distribution centred around 11300 people. This is not a surprise since the ward boundaries have been designed to  have approximatively the same population size.

In the top right panel of  Fig.~\ref{fig12}, we show instead the cumulative frequency distribution $P(s^{in}>s^*)$ for the number of people $s^{in}$ working in a given ward.
This is a skew distribution with a broad tail, well fitted by a power law $P(s^{in}>s^*)\propto s^{\gamma+1}$,   with exponent $\gamma=-2.24$. This reflects the fact that the approximation \textit{population size}/\textit{employment data}, that has been shown to be valid in the case of counties in US, is not valid in the case of London's wards. In particular, we see that employment  $s^{in}$ follow a distribution that suggests a complex and hierarchical organization for these resources within the city.

 In fact, from the bottom left panel of Fig.~\ref{fig12}, where we measure the average number of employees $<s^{in}(p)>$ as a function of the ward population size, we can see that apart from some non trivial deviations for small population size, there are not significant correlations. These deviations are related to the fact that the most significant employment locations in London often have a very small population.

We can now check whether Eq. 5 could be a more appropriate choice in order to describe commuting flows inside of a city, instead of Eq. 4. In the bottom right panel of Fig.~\ref{fig12}, we show the results of Eq. 5 applied to the commuting between GLA wards, versus the real commuting flows, in the same style and notation of Fig. ~\ref{fig6}. We notice that the plot is very similar to the one obtained using Eq. 4 and the $R^2=0.00$ tells us that using Eq.5 instead of Eq.4 does not improve the goodness of the fit. This implies that failure of the radiation model in forecasting urban commuting flows does not reside in approximating population size/employment, but in the complexity of the system \cite{gonz}.

\section{Conclusions}

Human mobility is an outstanding problem in science. In more than one century of active work and observation, the gravity model has been considered the best option to model such a phenomenon  \cite{ref6_wilson}. The appearance of a new statistical model based on physical science
\cite{ref1_simini} has re-opened the debate on the topic. In particular the apparent independence of the radiation model from metrical distance and its property of being parameter-free is a significant and desirable change from past practice. The model needs to be tested in many different circumstances so that its wider applicability can be assessed.

In this paper we address the reliability of the radiation model against the gravity model for large scale commuting and transportation networks in England and Wales and for the intra-urban commuting and transportation network for the London region.

The first thing we notice is that both models fail to
describe human mobility within London. In this sense we argue that
commuting at the city scale still lacks a valid model and that
further research is required to understand the mechanism behind
urban mobility. In fact, the phenomena of socio-geographical segregation
\cite{ref9_theil} and residential/business ward specialisation \cite{ref1_simini} are key drivers in determining the structure of flows and the density of population in the city and these are not reflected by these statistical models \cite{gonz}.

For England and Wales, we first introduce the correct normalisation for finite systems in the radiation model. Such a normalisation
affects the flows from London by a factor of 17\%.
Then we notice that the models are not very good in describing transportation data, such as bus and train flows, while they can be considered acceptable  for  modelling the commuting flows.
The gravity  model II of Eq.\ref{eq_2_} fails to describe commuting models, and confirms that commuting correlations with population at origin and destination is not just linear.
The gravity model is satisfactory in describing the commuting flows and surely much better than the radiation model, even if the latter has the advantage to be parameter free that turns out to be useful in the cases where there is no data available to estimate any parameters.

Nevertheless from the fluctuation analysis it emerges that there is a consistent portion of the distance/destination population phase-space where the radiation model gives better estimates of the gravity model in terms of S{\o}rensen-Dice coefficient.
This means that for large distances and small and moderate destination population scales, the principles of the radiation model are reliable and that mobility patterns can be approached by a diffusion model where intervening opportunities on the commuting paths prevail on the distance of such paths. However, the modest overall radiation model performance in terms of $R^2$ indicates that more research on the subject has to be done in order to improve the model reliability.

Other ways to represent the commuting system are possible. For example, if we were to grid all the data thereby strictly defining population and employment as density measures, this would change the dynamics of the gravity and radiation models in that they have been originally specified to deal with counts of activity data like population and employments rather than their densities. Moreover the tradition in this field is to work with data that is available in administrative units rather than approximate that data on a grid because these units reflect changes in the spatial system over time. We  believe that the best way to conduct this study is to consider urban conglomerations as the natural entities involved in commuting flows. This choice relates to a well settled tradition in statistical physics that consider cities as well defined entities, such as in the Zipf's and Gibrat's law.

\subsection*{Acknowledgements}
JS and APM were partially funded by the EPSRC SCALE project (EP/G057737/1) and MB by the ERC MECHANICITY Project (249393 ERC-2009-AdG).
Further, we would like to thank the anonymous reviewers for their constructive feedback that has improved the paper, especially the suggestion to use the S{\o}rensen-Dice coefficient as an alternative error metric.


\begin{thebibliography}{99}

\bibitem{ref1_simini} F. Simini, M. C. González, A. Maritan, A.L. Barabási,
A universal model for mobility and migration patterns, Nature 10856
(2012).

\bibitem{ref2_noulas} A. Noulas, S. Scellato, R. Lambiotte, M. Pontil, C.
Mascolo, A tale of many cities: universal patterns in human urban
mobility, PLoS ONE \textbf{7}, e37027 (2011).

\bibitem{carey} H.C. Carey,  Principles of Social Science, J. B. Lippincott \& Co., Philadelphia, PA, (1959).

\bibitem{ref3_zipf} G.K. Zipf,  The P1 P2/D hypothesis: On the intercity movement of
persons. American Sociological Review \textbf{11}, 677-686 (1946).

\bibitem{ref4_stouffer} S.A. Stouffer, Intervening opportunities: a theory relating
mobility and distance. American Sociological Review \textbf{5},
845-867 (1940).

\bibitem{ref5_ballerini} M. Ballerini, N. Cabibbo, R. Candelier, A. Cavagna, E.
Cisbani, I. Giardina, V. Lecomte, A. Orlandi, G. Parisi, A.
Procaccini, M. Viale, V. Zdravkovic, Interaction ruling animal
collective behavior depends on topological rather than metric
distance: Evidence from a field study, PNAS \textbf{105}, 1232
(2008).

\bibitem{ref6_wilson} A.G. Wilson, Entropy in urban and regional modelling, Pion
Limited, London (1970).

\bibitem{alonso} W. Alonso,  A Theory of Movements: Introduction, Working Paper No. 266, Institute of Urban and Regional Development, University of California, Berkeley, CA  (1976).

\bibitem{ref7_zipf49} G.K. Zipf, Human behaviour and the principle of least effort: An
introduction to human ecology, Addison-Wesley Press (1949).

\bibitem{dorog} S.N. Dorogovtsev, J.F.F. Mendes, Evolution of Networks: From Biological
Nets to the Internet and WWW, Oxford: Oxford University Press (2003).

\bibitem{ref8_web} http://www.neighbourhood.statistics.gov.uk/ (last visited
25-06-12).

\bibitem{elsa} E. Arcaute, E. Hatna, P. Ferguson, H. Youn, A. Johansson, M. Batty, City boundaries and the universality of scaling laws,  arXiv:1301.1674, to appear (2013).





%\bibitem{ref10_masucci} A.P. Masucci, G.J. Rodgers, The Network of Commuters in London,
%Physica A \textbf{387}, 3781 (2008).

\bibitem{sorensen}
T.A. S{\o}rensen (1948) A method of establishing groups of equal amplitude in plant sociology based on similarity of species and its application to analyses of the vegetation on Danish commons, Biol. Skr., \textbf{5} (1948), pp. 1-34

\bibitem{dice} L.R. Dice (1945) Measures of the Amount of Ecologic Association Between Species, Ecology \textbf{26} (3): 297–302.

\bibitem{gonz} P. Wang, T. Hunter, A.M. Bayen, K. Schechtner, M.C. Gonz\'alez, Understanding Road Usage Patterns in Urban Areas, Scientific Reports \textbf{2},1001 (2012). 

\bibitem{ref9_theil} H. Theil, A.J. Finizza, A Note on the Measurement of Racial
Integration in Schools, Journal of Mathematical Sociology
\textbf{1}, 187 (1971).


\end{thebibliography}
\end{document}